\documentclass[twocolumn,preprintnumbers,amsmath,amssymb,prb,lettersizepaper]{revtex4}
\usepackage{layout,float}
\usepackage{amsmath,amssymb}
\usepackage{graphicx}
\usepackage{dcolumn}
\usepackage{bm}

\DeclareGraphicsExtensions{.bmp}

\newcommand{\be}{\begin{equation}}
\newcommand{\ee}{\end{equation}}
\newcommand{\fr}{\frac}

\newcommand{\iz}{\left}
\newcommand{\de}{\right}

\newcommand{\tb}{\textbf}
\newcommand{\bit}{\begin{itemize}}
\newcommand{\eit}{\end{itemize}}

\newcommand{\bra}[1]{\langle #1|}
\newcommand{\ket}[1]{|#1\rangle}

\begin{document}
\title{\large{SIZE CHANGE EFFECT ON THE OPTICAL BEHAVIOR OF ULTRA SMALL METAL PARTICLES}}
\author{ Mario Zapata Herrera}
\author{Angela S. Camacho B.}

\affiliation{Departamento de F\'isica, Universidad de los Andes}
\email{m.zapata62@uniandes.edu.co}
\date{\today}
	
\begin{abstract}
In this paper we analyse the importance of a detailed description of the electronic transitions in ultra-small nanoparticles through the optical response to very small  changes of size in systems, whose dimensions are in the subnanometric scale.
We present a carefully calculation of the optical response of these systems  by using the exact eigen-energies and wave functions for nanospheres with diameter smaller than $10$nm, to obtain the dielectric function under different conditions of confinement. The aim is to use the so obtained dielectric function to present the absorption spectra of one electron confined in a sphere in two cases: 1) infinite confinement and 2) finite confinement, in which, the value of the wells depth is carefully calculated after adjusting the number of atoms that composed each sphere, so that the energies and dipole matrix elements give a more accurate information of the optical response. Moreover, we extend the calculation of this dielectric function for obtaining the optical constants needed to find the plasmon frequency, throught a numerical method of finite elements that solves the Maxwell’s equations, in order to obtain the enhancement of the near electric field. We show an interesting behavior for particles sizes less than $10$ nm, finding that the variation induced in the eigen-energies, through slight size changes in the particle, provides significant variations in the optical response of these nanoparticles. This effect, can be observed in the optical absorption spectra and in the localized surface plasmon energies as confirmed in recent observation of plasmonic phenomena at the subnanometer to atomic scale.
\end{abstract}

\maketitle
\section{Introduction}
In the past few years, attention has focused on nanoparticles composed of noble metals, because they show localized  excitations near the surface that are originated from the free electrons of the metal,\cite{Puska,Maier,Ritchie,Ekardt} which are collective oscillations of the conduction band electrons in the terahertz regime \cite{THz1,THz2,THz3}, so-called \textit{plasmons}. Since the possible applications in this range are extremely interesting, such as cancer therapy\cite{Cancer1}, nanophotonic devices \cite{Nanofotonics1,Stockman1,Nanophotonics2}, biosensing\cite{biosensing} up to  catalysis \cite{catalisis1,catalisis2}, there has been considerable progress understanding the underlying physics through experimental and theoretical research of small particles properties. \cite{Townsend,Willets,Juluri} Particularly, the collective excitations known as localized surface plasmons (LSP) are subject of analysis because they enable strong optical absorption and scattering in subwavelength structures. These properties depend on the nanoparticle’s material, size, shape, and the refraction index of the sorrounding material \cite{Juan,Carsten,Wei,Murray}. Each one of these parameters is worth of being studied in order to understand the optical behavior at the nano-scale.\\

Most of the calculations assume the Mie model\cite{Mie} because it shows good agreement with experiments in nanoparticles up to $10$nm size\cite{JandC}. For smaller particles, this model achieves poor agreement, and in particular, when dealing with particles in the size-range of $1-10$nm,  the surface starts to become more and more important compared to the bulk response and thus influences heavily the collective modes localized at the surface region. On the other hand, the quantum effects should be significant\cite{Garcia} due to the electrons could be able to escape of the particle under finite confinement, and new interesting effects appear for these systems that are in the subnanometric sizes. Plasmon oscillations and spill out could balance tourning off the optical response of the particles, as proponed recently Monreal \textit{et al}, \cite{Monreal} where they examine electron spill-out effects, which can move the surface plasma energy both toward the red or the blue and they can be comparable to or even stronger than the quantum size effect. However, their model remains classical in spite of  introducing a semiconductor like gap at the Fermi level. This model explains the experimental reported red and blue shifts of the plasmon frequency in very small nanoparticles, but the subnanometric sizes open the door to understand them as a pure quantum effect.\\

Following this idea we propose to analyze the effects of quantization by presenting an exact calculation of the dielectric function, compared with less rigurous approaches and with accurate experiments in order to examine the optical properties of systems in the limit of subnanometric size scale. We focus this work on the quantum description of optical response of ultra-small spherical nanoparticles, whose diameters are less than $10$nm. We describe the dielectric function of these systems starting from the assumption that one electron confined in the nanoparticle only can have discretized energies and we make carefully calculations of the allowed electron energies and the corresponding wave-functions for  infinite and finite confinement cases as function of the particle size (diameter). In both situations we use the analytical complete eigen-functions to obtain the dielectric function, which give us confidence in the description of the optical response. We test our results of the dielectric function through optical absorption and enhancement field factor,  which allow us to compare with recent experimental results\cite{Scholl}. Starting from infinite spherical confinement with  asymptotic values for energies and wave functions (Model A) and exact calculation of both energies and wave functions (Model B), we examine the  effect of the exact values on the dielectric function for ultra-small particles. These results are also compared with finite spherical  wells confinement, whose depth is obtained from the number of atoms in each particle (Model C).\\

This paper is organized as following,  in Sec. \ref{<Dielectric_F>}, we present the main steps to obtain analytically expressions for the dielectric function within three approximations: 1) infinite confinement as in ref.\cite{Scholl} using approximate energies and asymptotical wave-function,  which is labelled as model A in the rest of the paper; 2) infinite confinement, with  exact energies and wave functions for infinite spherical well, as Model B,  and 3) finite confinement using the exact energies but the wave-function of the infinite well, as Model C. Sec. \ref{<absorption_S>} is dedicated to the optical absorption of one particle confined in an spherical well, where we can compare our results with experiments, and discuss the differences in absorption coming only from quantum confinement, when the electrons have to stay within the particle and when they have a probability of being outside depending on the particle size with the aim of seeing the effect on the optical response of these ultra small particles in a wide energy range. In Sec. \ref{<EF>}, we show a reproducible behavior of the plasmon energy depending on confinement, and finally, in Sec. \ref{<Conclusiones>}, we compare our results with experiments and the theoretical model based on the spill out effect.
\section{\label{<Dielectric_F>}The dielectric function}
When the size of a particle becomes smaller than $10$nm, the continuous electronic band of a nanoparticle breaks up into observable discrete states. Such effects have been observed in experiments with metallic nano-particles, \cite{Genzel,Scholl}  showing a  notably difference between the results obtained in the $10-100$ nm regime  and the ‘quantum regime’ (particles below $10$ nm of radius). \\

As a first approximation, to describe the optical response of an electron gas in a metal nanoparticle,  Genzel and Martin \cite{Genzel} showed a simplified quantum model where  the free electron gas is confined in a cubic potential well with infinite walls, where the dielectric function of a single metal nanoparticle, under the influence of an electromagnetic wave with  frequency $\omega$ and z-polarization,  is given by 
\be
\epsilon(\omega)=\epsilon_{\infty}+\fr{\omega_p^2}{N}\sum_{i,f}\fr{s_{if}(F_i-F_f)}{\omega_{if}^2-\omega^2-i\omega \gamma_{if}},
\label{eq1}
\ee
with the terms $\epsilon_{\infty}$, $s_{if}$, $\omega_{if}$ and $\gamma_{if}$ are respectively, the interband contribution of core electrons, the oscillator strenght, the frequencies and the damping for the dipole transition from an initial state $i$ to another final state $f$; and $F_i$ and $F_f$  are the values of the Fermi-Dirac distribution function for the initial and final states. The oscillator strenght $s_{if}$  corresponds to the dipolar form in the z-direction in terms of the initial $ |i\rangle$ and final $ |f\rangle$ states
\be
s_{if}=\fr{2m\omega_{if}}{\hbar}|\langle f|z|i \rangle |^2.
\label{strenght}
\ee
where the parameter $m=M/N$ is given in terms of the mass of electron $M$ and the total number of confined atoms $N$. Using different types of confinements, we calculate the eigen functions and energies, within the A, B and C models described before.\cite{parametros}

\subsection{\label{<SchollModel>}Free electron gas in a spherical infinite well confinement: Exact solution}
We consider an infinite spherical potential well to solve the Schroedinger equation for one electron,  with the appropiate boundary conditions  finding  a set of wavefunctions of the form \cite{Arfken} 
\begin{equation}
\psi(r,\theta,\phi)=\fr{1}{|j_{l+1}(\alpha_{nl})|} \sqrt{ \fr{2}{R^3}}  j_l \left( \fr{\alpha_{nl}}{R}r \right)Y_l^m(\theta,\phi),
\label{Psi}
\end{equation}
where $j_l$ represents the spherical Bessel functions, $Y_l^m$ the standard spherical harmonics, and $\alpha_{nl}$ is the $n$-th zero of $j_l$. The eigen-energies $E_{n,l}$ of this problem in a quantized form associated to the $l$-th spherical Bessel function are given by 
\begin{equation}
E_{n,l}=\frac{\hbar^2\alpha_{ln}^2}{2MR^2}.
\label{E}
\end{equation}
The oscillator strenght in eq. (\ref{strenght}) can be calculated, taking $z=r \cos\theta$ in the expression
\be
\begin{aligned}
\begin{split}
|\bra{f}z\ket{i}|=&\int_{0}^{2\pi}\int_{0}^{\pi}\int_{0}^{R} dr d\theta d\phi r^3 \sin\theta \cos \theta \\
& \times\Psi_{n_f,l_f,m_f}^*(r,\theta,\phi) \Psi_{n_i,l_i,m_i}(r,\theta,\phi),
\end{split}
\end{aligned}
\ee
where the angular integral has the form
\be
\begin{aligned}
\begin{split}
I_{ang}=&\sqrt{\fr{(l_i+m_i+1)(l_i-m_i+1)}{(2l_i+1)(2l_i+3)}}\delta_{\Delta l,+1}\\
&+\sqrt{\fr{(l_i+m_i)(l_i-m_i)}{(2l_i+1)(2l_i-1)}}\delta_{\Delta l,-1}
\end{split}
\end{aligned}
\label{angular}
\ee
and, the radial term,
\be
\begin{aligned}
\begin{split}
I_{rad}=&\fr{1}{|j_{l_f+1}(\alpha_{n_fl_f})|} \fr{1}{|j_{l_i+1}(\alpha_{n_il_i})|}\left( \fr{2}{R^3}\right)\\
&\times\int_{0}^{R}dr  j_{l_f} \left( \fr{\alpha_{n_fl_f}}{R}r \right)r^3  j_{l_i} \left( \fr{\alpha_{n_il_i}}{R}r \right).
\end{split}
\end{aligned}
\ee
Then, the integral to calculate is simply
\be
|\bra{f}z\ket{i}|=I_{ang}\times I_{rad} 
\ee
which  is different to zero only for values  $\Delta l=l_f-l_i=\pm 1$. This description corresponds to our labelled Model B.
\subsection{Free electron gas in a spherical infinite well confinement: Approximate solution}
One way to obtain a simplified solution of the problem is to use the approximation\cite{Scholl}
\begin{align}
E_{n,l}&=\frac{\hbar^2\pi^2}{8MR^2}(2n+l+2)^2\notag\\
&=\frac{\hbar^2}{2MR^2}\left[\pi\left(n+\frac{l}{2}+1\right)\right]^2,\label{ES}
\end{align}
for the eigen-energies, and  the asymptotic form for the  wavefunctions\cite{Arfken},
\begin{equation}
j_l(x)\approx \frac{1}{x}\cos\left[x-\frac{\pi}{2}(l+1)\right],
\label{approx}
\end{equation}
valid only for $x\gg l^2/2+l$.\\
\begin{table}[htdp]
\begin{center}
\begin{tabular}{|m{0.8cm} |m{1.3cm} |m{1.3cm}|m{1.3cm} | m{1.3cm} |m{1.3cm}| }\hline
& $n=0$ & $n=1$ & $n=2$ & $n=3$ & $n=4$\\\hline
$l=0$ & 376.03 & 1504.11 & 3384.25 & 6016.45 & 9400.71\\
& 376.03 & 1504.11 & 3384.25 & 6016.45 & 9400.71\\\hline
$l=1$ & 769.26 & 2273.77 & 4530.04 & 7538.31 & 1129.86\\
& 846.64 & 2350.18 & 4606.35 & 7614.57 & 11374.9 \\\hline
$l=2$ & 1265.57 & 3151.57 & 5785.61 & 9170.69 & 13307.4 \\
& 1504.11 & 3384.25 & 6016.45 & 9400.71 & 13537.0\\\hline
$l=3$ & 1860.45 & 4134.43 & 7148.86 & 10912.1 & 15426.0\\
& 2350.18 & 4606.35 & 7614.57 & 11374.9 & 15887.2\\\hline
$l=4$ & 2550.93 & 5219.83 & 8617.81 & 12760.9 & 17653.2\\
& 3384.25 & 6016.45 & 9400.71 & 13537.0 & 18425.4\\\hline
\end{tabular}
\end{center}
\caption{Comparison between energies given by eq.(\ref{E}) upper and by eq.(\ref{ES}) lower in units of meV when $R=1$ nm for various values of $l$ and $n$.}
\label{beta}
\end{table}

\begin{figure}
\begin{center}
\includegraphics[scale=0.32]{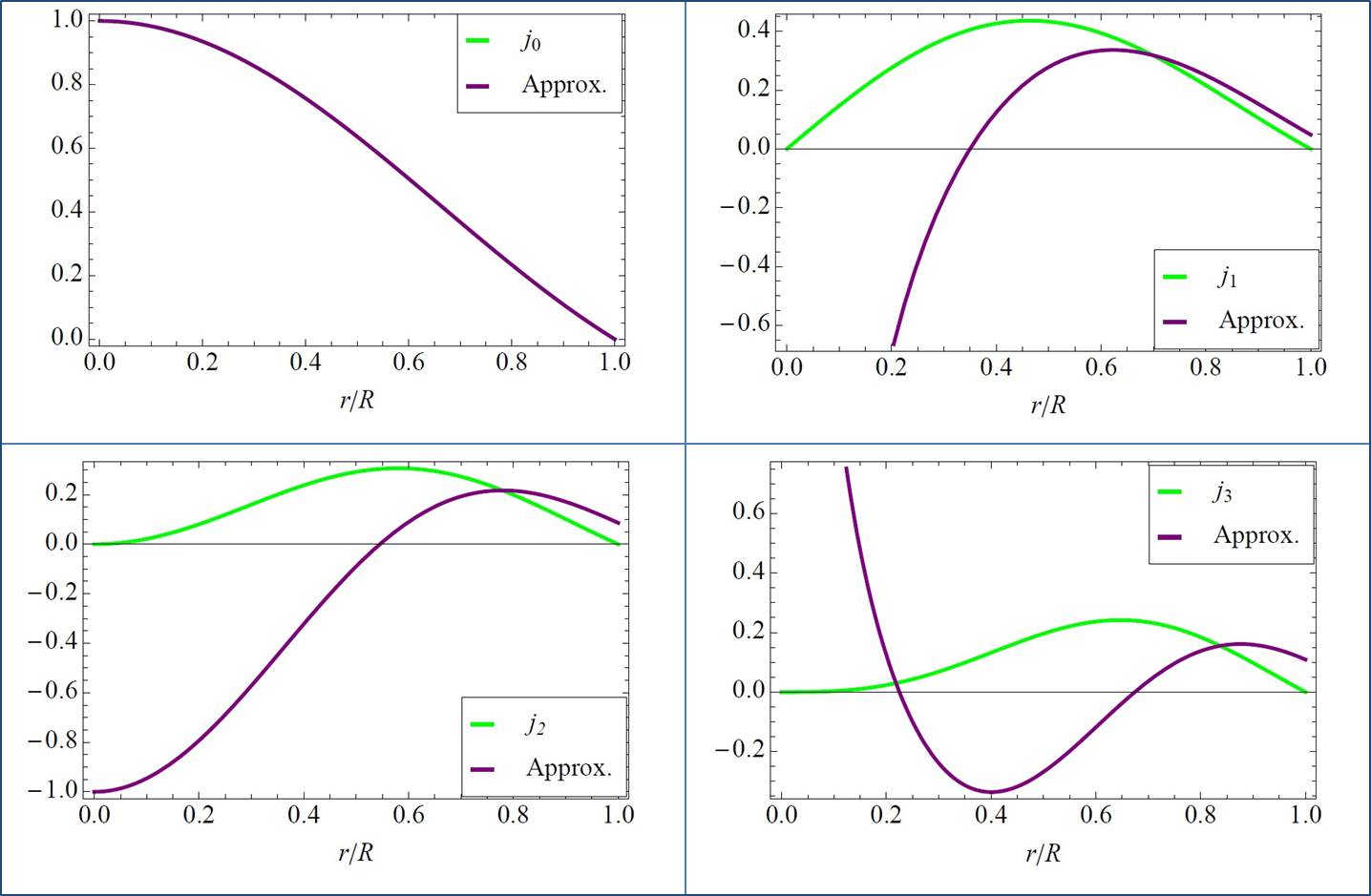}
\caption{Comparison between $j_l$ and its asymptotic approximation (\ref{approx}) versus $r/R$ for different $l$'s and $n=1$.}
\label{j0}
\end{center}
\end{figure}

In Table \ref{beta}, we compare the energies given by eq. (\ref{E}) (Model B) and eq. (\ref{ES})  (Model A) for nanoparticles with $R=1$nm,  where we observe a good agreement only for $l=0$, and a significant overestimation for $l >0$. Besides this, in Figure \ref{j0} we show $j_l(kr)=j_l(\alpha_{ln}r/R)$ and its asymptotic approximation given by eq. (\ref{approx}) versus $r/R$ and $n=1$. As we can see, the approximate and the exact Bessel functions are identical for $l=0$, but they differ remarkably for the  $l=1, 2$, and $3$ quantum numbers. Therefore, we conclude that the use of exact wave functions (Model A) are needed to give an accurate description of the optical response in these systems. 
\subsection{Quantum finite confinement}
\begin{figure}[H]
\begin{center}
\includegraphics[scale=0.33]{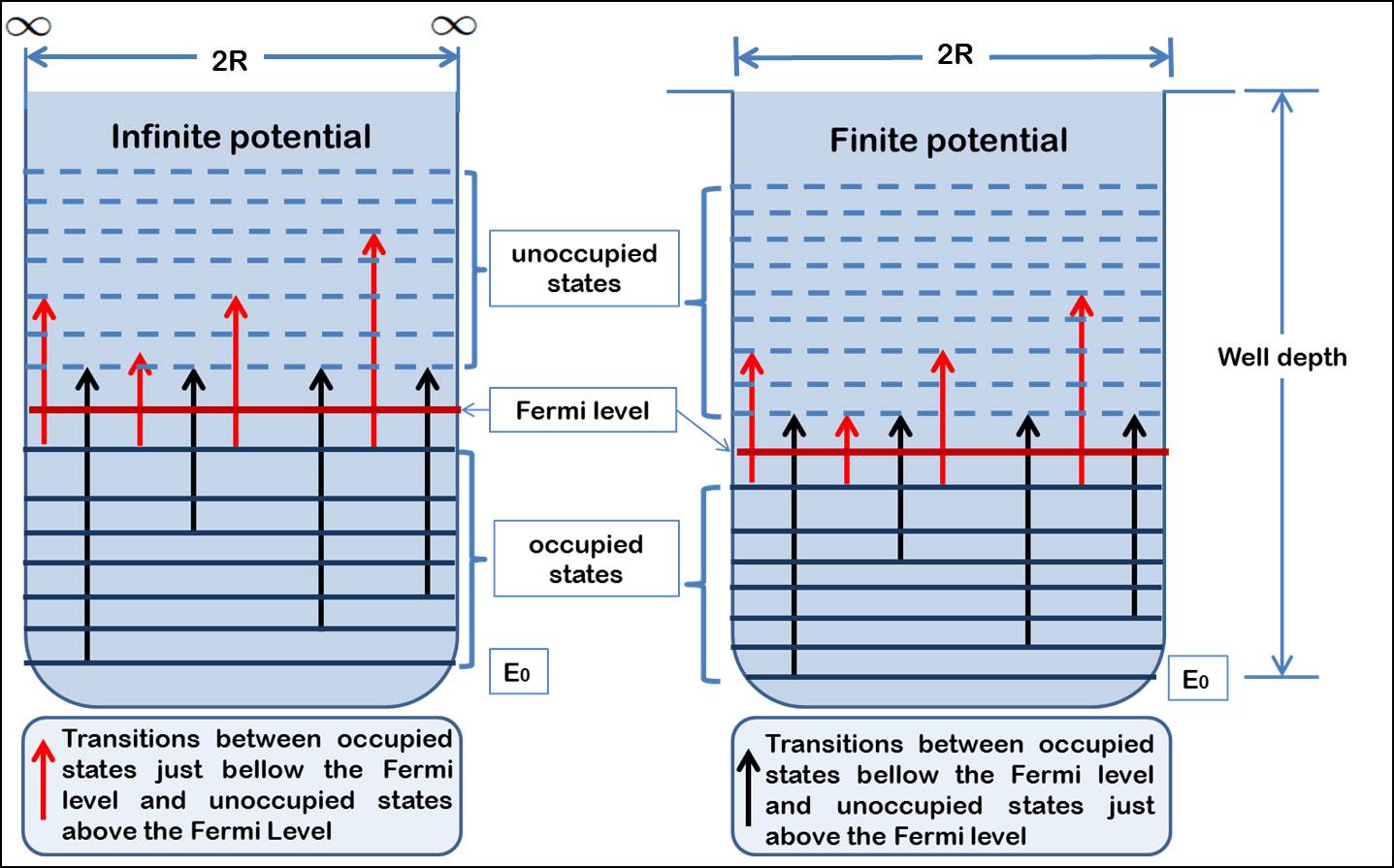}
\caption{\normalsize{Schematic diagram illustrating quantum infinite  potential and quantum finite potential, to evaluate the dipolar contributions in the dielectric function. Here we show some transitions  when the energy levels are quantized. }}
\label{diagrama-finito}
\end{center}
\end{figure}

Going a step further, we consider a finite spherical confinement, which is related  with the number of atoms contained in a spherical nanoparticle. We fit a relation derived of  \textit{ab initio} calculations made by He \textit{et. al.}\cite{Yi}, by taking the number of particles as function of radius as
\be
N(R)\approx 246R^3.
\ee
Based on this relation, then we calculate the last occupied electronic state following the Pauli exclusion principle. This energy value is defined as the Fermi level and  varies  with the nanoparticle radius. To establish the depth of the well, we consider the work function $W(R)$ for small metallic nanoparticles following ref.\cite{Wood}, which is defined as the energy required to remove one electron from the nanoparticle, which is given by
\be
W(R)=4.37+\fr{5.4}{R(\AA)},
\ee
where the $4.37$ value is the average work function  for Bulk Ag (in eV) and $5.4=\fr{3}{8}e^2$ is a parameter that accounts for the difference in the work function for a conducting plane and a sphere. \cite{Wood} Considering this value, we define the depth well $D_W$ as the sum of the Fermi energy level $FE$ and $W(R)$ for each nanoparticle radius, which reads as
\be
D_W(R)=-(FE(R)+W(R)).
\ee
Using this result, we solve the Schroedinger equation numerically, calculating the eigenvalues for each nanoparticle. In order to obtain the corresponding dielectric function, we need the oscillator strenght for the allowed transitions using the wave functions of infinite confinement, which are needed to obtain the dielectric function as can be seen in eq. (\ref{strenght}). We choose transitions between occupied  states just below  the Fermi  level  and  unoccupied states above the Fermi level, and transitions between the unoccupied  state just above the Fermi level and occupied states below the Fermi level, in a wide range ($0-6$eV), following the selection rule $\Delta l=\pm 1$, which gives us a complete information of the allowed electronic transitions in our systems. We show in Fig. \ref{diagrama-finito}, a schematic diagram that illustrates this assumptions for quantum B and C Models.
\subsection{Confinement effects on the optical response}
\begin{figure}
\begin{center}
\includegraphics[scale=0.04]{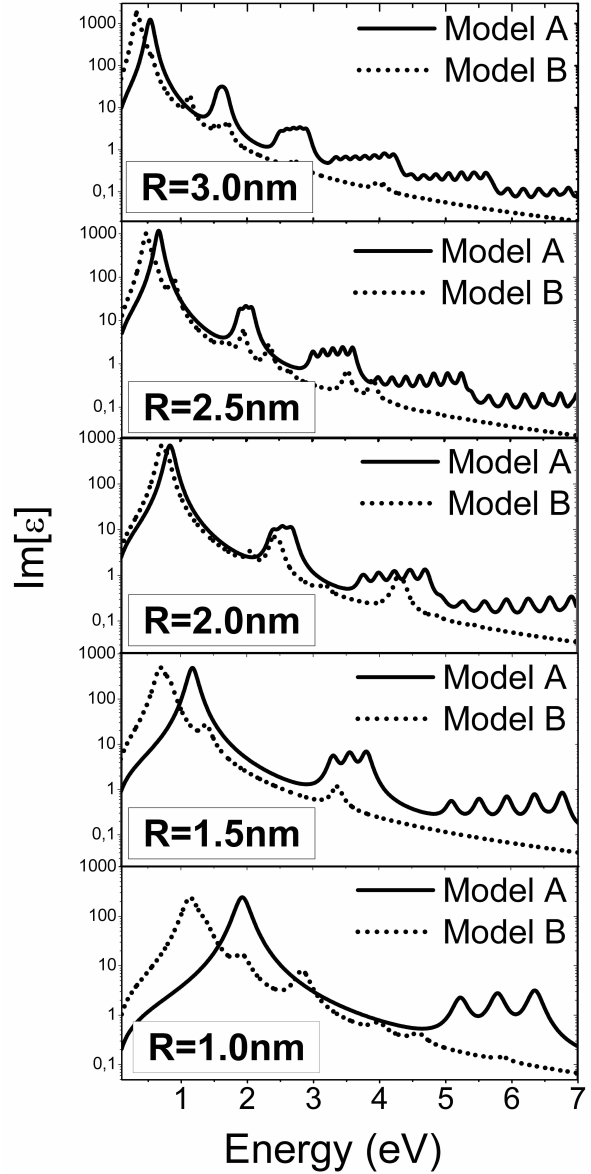}
\caption{\normalsize{Imaginary  part of the dielectric function for a silver nanoparticle as function of radius  based on  Model A (continuous lines) and  Model B (dashed lines).}}
\label{dielectrica}
\end{center}
\end{figure}

Now, we calculate the dielectric function based on  eq. (\ref{eq1}), and  present the  imaginary part of the dielectric function for several nanoparticle sizes in Fig. \ref{dielectrica}.  The continuous  black lines, corresponding to the Model A, show besides the main peak, several possible transition with smaller intensity, while the dashed lines, Model B,  reduces significantly the number of transitions in the region of interest, as an effect of the oscillator strenght. On the other hand, the energy position of the main peaks is always blue-shifted  with respect of Model B for each radius.\\

\begin{figure}
\begin{center}
\includegraphics[scale=0.33]{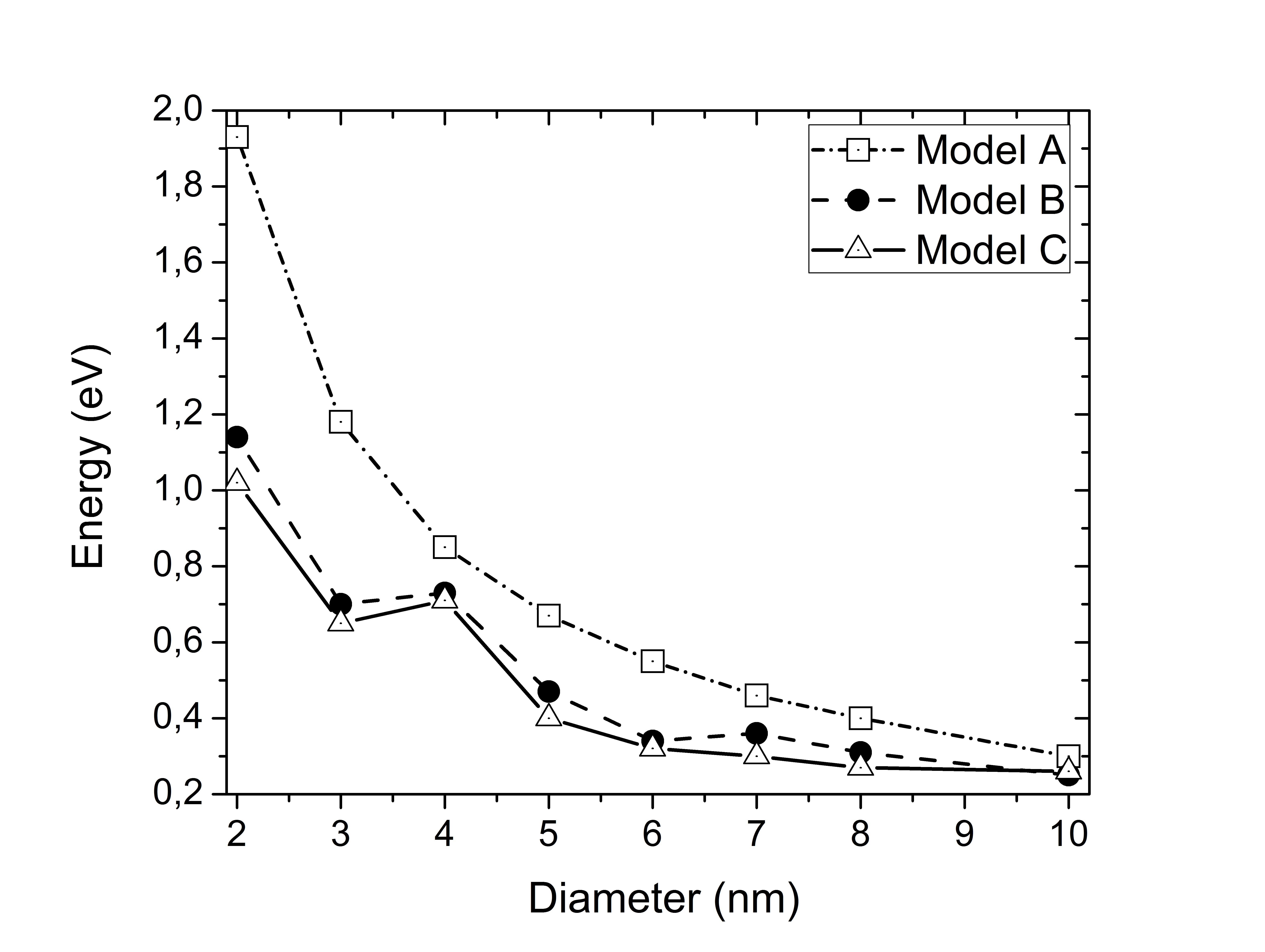}
\caption{\normalsize{Main peak in the imaginary part of the dielectric function for each model used. Squares represent the Model A,  black points Model B and triangles Model C. }}
\label{imaginariadielectrica}
\end{center}
\end{figure}

With the aim of examining the trend of the peaks of energy in the imaginary part of the dielectric function as function of radius and to compare the results of the three models we present Fig. \ref{imaginariadielectrica}. While model A predicts a monotone blue-shift as the particle radius  decreases, whose value achieves almost $2$eV for $R=1$nm,  Models B and C, show a maximal blue-shift about $1$eV. However, the shifts between consecutive diameters oscillate.  
\section{\label{<absorption_S>}Absorption spectra}
\begin{figure}[H]
\begin{center}
\includegraphics[scale=0.34]{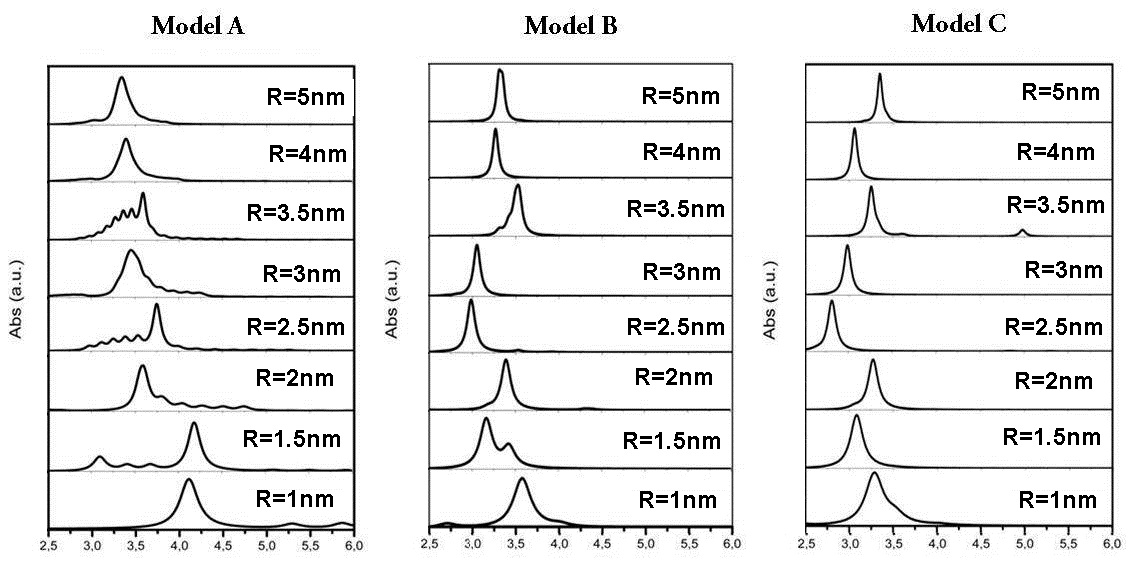}
\caption{\normalsize{Absoption spectra for several particle radius, using Model A (left), Model B (center) and Model C (right).}}
\label{Absorption}
\end{center}
\end{figure}

In this section, we compare the absorption peaks following the three approaches described above. To calculate the absorption spectra, we use the expression
\be
\sigma_{ext}(\omega)=\fr{9f\omega \epsilon_m^{3/2}\Im[\epsilon(\omega)]}{c(\Re[\epsilon(\omega)]+2\epsilon_m)^2+(\Im[\epsilon(\omega)])^2},
\ee
where $f$ is a fraction of volume in the media, $\epsilon_m$ corresponds to the surrounding media and $c$ the light speed.\cite{constantes} For particles of radius $5$nm there is no appreciable  difference between the three calculations (see Fig. \ref{Absorption}), showing  all of them a peak at $3.3$ eV,  while for smaller particles we obtain both red and blue shifts that are coming from the confinement type. Looking from the top to the bottom of Fig. \ref{Absorption}, the particle radius decreases and Model A (at the left), show a rather smooth blue-shift while Models B and C (center and right), present red and blue shifts alterning as function of the nanoparticle radius.
We observe oscillations of the maximal absorption for nanoparticles smaller than $10$nm ($R=5$nm). The high dispersion of the experimental EELS measurements\cite{Scholl} in this region also shows big oscillations for particles smaller than $10$nm of diameter. Our predictions are located below the region of experimental error, which can be corrected through the $\epsilon_{\infty}$ parameter. The three approximations are in good  agreement to each other and to the experiment, showing changes of red and blue shifts at the same particle sizes. \\
With the aim of testing our three models of dielectric function related to the localized surface plasmon resonance in silver ultra small nanoparticles, we present in the next section the enhancement field factor. 
\section{\label{<EF>}Enhancement of the near field}
The dielectric function $\epsilon(\omega)=\Re[\epsilon(\omega)] + i\Im[\epsilon(\omega)]=\epsilon_1(\omega)+i\epsilon_2(\omega)$ following the three models,  allows to obtain three different sets of optical constants:  refractive index $\eta$ (related with the polarization), and the extinction coeficient $\kappa$ (related with the optical absorption) through the equations 
\be
\eta^2=\fr{\epsilon_1}{2}+\fr{1}{2}\sqrt{\epsilon_1^2+\epsilon_2^2}\label{eta}
\ee
and
\be
\kappa=\fr{\epsilon_2}{2n}\label{kappa}.
\ee
These two quantities contain all the optical information needed in the numerical solution of the Maxwell equations,  in order to predict the enhancement of the near field as the particle radius is decreasing. To solve the problem of light-nanoparticle interaction, we solve the Maxwell equations  respect to the intensity of the scattered electric field vector $\vec{E}$, which is given by 
\be
\nabla \times \mu_r\iz(\nabla \times \vec{E}\de)-k_0^2\iz(\epsilon_r-\fr{i\sigma}{\omega\epsilon_0}\de)\vec{E}=0,\label{Maxwell}
\ee 
where $\mu_r$ is the relative permeability (taken as $=1$ for metallic systems), $\epsilon_r$ the complex relative permitivity, $\sigma$ is the nanoparticle conductivity  and $k_0=\omega \sqrt{\epsilon_0\mu_0}=\fr{\omega}{c_0}$, with $c_0$, $\epsilon_0$ and $\mu_0$ the corresponding values of light velocity, the permeability and permitivity in vacuum, respectively. \\

We calculate expressions for the dielectric function dependent of the incident electromagnetic wave frequency $\epsilon_r(\omega)$ as input parameters, which are  calculated  for each particle size and confinement potential, to solve the vector Maxwell equations numerically, by means of a standar finite elements method (FEM) using the COMSOL Multiphysics package. We define the enhancement field factor (EFF) as the norm of the ratio between the scatter electric near field just in the north pole of the spherical nanoparticle  $\vec{E}_{out}$ and an incident $z$-polarized electric field $\vec{E}_{inc}$.
 \be
 EFF=\fr{|\vec{E}_{out}|^2}{|\vec{E}_{inc}|^2}
 \ee
The obtained resonances give the frequencies of the surface local plasmon, and we observe that we obtain values located at the same energies as the absorption peaks, which follow the same behavior as discussed earlier.\\

\begin{figure}[h]
\begin{center}
\includegraphics[scale=0.06]{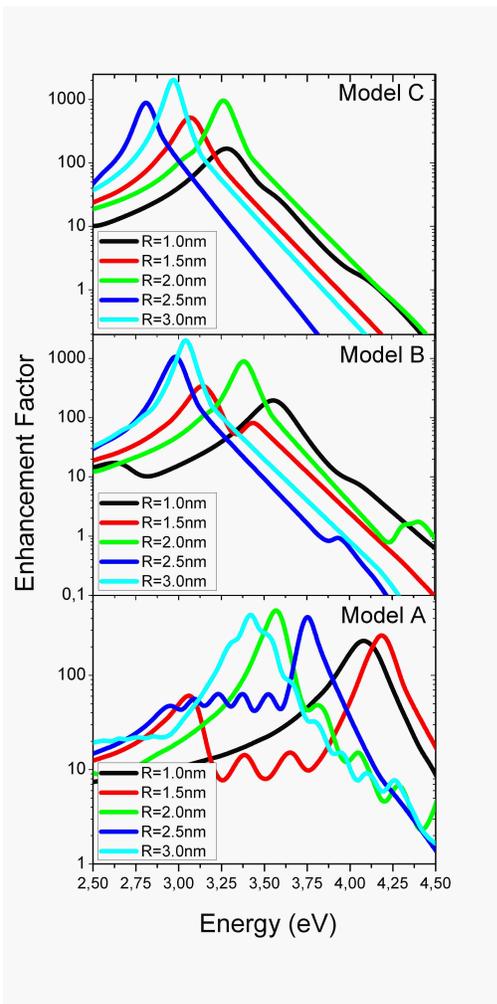}
\caption{\normalsize{Enhancement Field  Factor for a silver nanoparticle as function of radius  based on quantum approximate model (model A) \cite{Scholl}, infinite exact model (model B) and finite confinement (model C) of one electron in an spherical particle.}}
\label{factordeaumento}
\end{center}
\end{figure}

\begin{figure}[h]
\begin{center}
\includegraphics[scale=0.037]{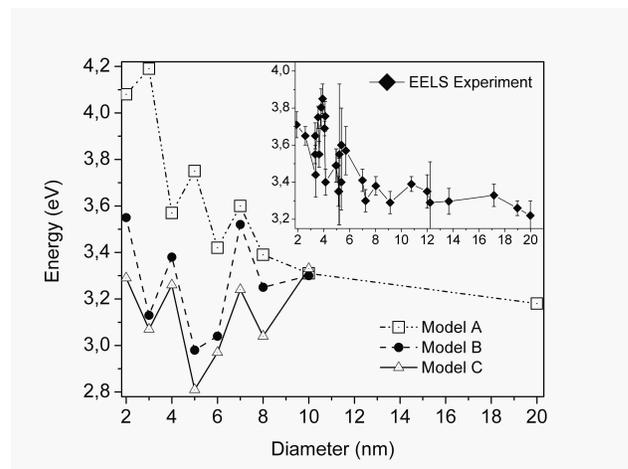}
\caption{\normalsize{Plasmon energy as a function of the spherical nanoparticles diameter. Squares, dots and triangles show the results when Model A, B and C are used. The inset present the experimental EELS results obtained by Scholl \textit{et. al.} in \cite{Scholl}.}}
\label{comparadas}
\end{center}
\end{figure}

If we look at the field factor for each particle radius in Fig. \ref{factordeaumento} from bottom to top, we observe a red-shift for all particle radius as a whole, depending on the used model. Comparing Model A with Model B, the change is due to the dipole matrix elements (oscillator strenght), while the differences between Model A and Model C are due to transition energies. Both effects show lower plasmon energies than the approximate Model A, mantaining the same parameters in the three models. However, they show an increasing in the enhancement field factor of about 4 times (compared with model A). The behaviour of the plasmon frequency within each model, present oscillations, that means red and blue shifts alternating. 
Finally we present in Fig. \ref{comparadas}, another feature that is interesting comparing with EELS experiment, which is the energy width of the region where the oscillations are detected. In Model A, the width achieves more than $1.5$eV, while for model B and C is only of $1$eV. On the other side, the experimental measurements show values ranging from $0.8$eV, with maximal error bars for $5$ and $3$nm diameters, where small changes in the particle size can give strong shifts, which depend on the sensitivity to the well depth. It means to the eigen energies and corresponding oscillator strenghts. As the particle decreases from $10$nm, the enhancement field factor decreases, indicating a weaker near field because the number of electrons that can resonate are getting less and less. However, for this range of nanoparticles the ratio surface/volume increases considerably, giving rise to strong dependence on the energy differences and oscillator strength between them.\\

In Fig. \ref{comparadas}, we show the localized surface plasmon resonance frequency, obtained by the maximum of enhancement field factor for each particle size and we observe that for particles smaller than $10$nm our three models provide the same behaviour, showing alternating red and blue shifts as the particles size decreases. In the inset we present the experimental EELS reported for the plasmon frequency in the same sizes range obtained by Scholl \textit{et. al.}\cite{Scholl},in which we can observe error bars of about one half electronvolts in the energy  plasmon spectrum for nanoparticles about $5$nm and  $3$nm. However, our three models give plasmon energy values for several particle sizes, showing in this spectrum, a kind of oscillations, as the particle size decreases. 
The confinement effect only is important for very small nanoparticles, as expected. In a recent paper \cite{Monreal} is discussed the screening effect on the surface plasmon compared to quantum size effect for ultra small nanoparticles, less than $10$nm. They show a monotone behavior, as result of a balance between size effect (blue shift) and spill out effect (red shift) based on pure classical  calculation. However, when comparing with EELS experiments,  the big error bars in the region of interest screen detailed optical response in this region. Our claim is that there are clear oscillations within the error bars that can be understood as confinements effects due to the energy differences and their allowed transitions, which are extremely sensitive to slightly changes in the nanoparticle size.
\section{\label{<Conclusiones>}Concluding Remarks}
We examine the optical response of ultra-small metal nanoparticles varying their sizes from $10$ to $2$ nm diameter in steps of 0,5 nm in order to observe the sensitivity to slight particle size difference. The reason for the oscillations in optical response is that very small changes in the energy spectrum induce significant changes in the allowed electronic transitions. We also compare two types of confinement,  using two sets of wave functions and we conclude that red shifts in the plasmon energies are mainly due to the oscillator strenghts, while the eigen-energies for finite and infinite confinement taken from the exact zeros of the Bessel functions (Models B and C) do not influence much the plasmon energies. But both of them differ considerably from the calculation based on asymptotic values (Model A).\\

We have shown that exact quantum calculations, based on optical response of one electron confined in a sphere, open a way to understand the optical properties of ultra-small nano-particles, which is to be completed with many body effects that surely appear in this region of particle sizes that limits the nanoscale and enters in the atomic scale.\\

In the region beyond nanometric scale, we could find optical response of nanoparticles strongly dependent on the change of the size due the changes of the allowed electronic transitions, within the discrete energy levels that should be taken into account carefully to understand the plasmon behavior in ultra small nanoparticles that are compossed of hundreds of atoms.\\

Other effects that should be taken into account to understand the optical response of ultra-small nano  particles, undoubtely require microscopic accurate description such as non local and non lineal effects.\\

\vspace{-1.25cm}
\bibliography{References}

\begin{thebibliography}{9}

\bibitem{Puska}{ M. J. Puska, R. M. Nieminem and M. Manninen, \textit{Electronic polarizability of small metal spheres}, Phys. Rev. B. \tb{31}, 6 (1985).}
\bibitem{Maier}{S. A. Maier, \textit{Plasmonics: Fundamentals and applications}. Springer, 2007}.
\bibitem{Ritchie}{R. H. Ritchie, \textit{Plasma losses by fast electrons in thin films}, Phys. Rev. \tb{106}, 5 (1957).}
\bibitem{Ekardt}{W. Ekardt, \textit{Work function of small metal particles: Self-consistent spherical jellium-background model}, Phys. Rev. B \tb{29}, 1558 (1984).}
\bibitem{THz1}{S. A. Maier, S. R. Andrews, L. Mart\'in-Moreno, and F. J. Garc\'ia-Vidal, \textit{Terahertz Surface Plasmon-Polariton Propagation and Focusing on Periodically Corrugated Metal Wires}, Phys Rev. Lett. \tb{97}, 176805 (2006).}
\bibitem{THz2}{Z. Li, Y. Ma, R. Huang, R. Singh, J. Gu, Z. Tian, J. Han, and W. Zhang, \textit{Manipulating the plasmon-induced transparency in terahertz metamaterials}, Opt. Express, \tb{19}, 8912 (2011).}
\bibitem{THz3}{M. Dragoman and D. Dragoman, \textit{Plasmonics: Applications to nanoscale terahertz and optical devices}, Progress in Quantum Electronics, \tb{32}, 1-41 (2008).}
\bibitem{Cancer1}{X. Huang, P. K. Jain, I. H. El-Sayed and M. A. El-Sayed, \textit{Plasmonic photothermal therapy (PPTT) using gold nanoparticles}, Lasers. Med. Sci. \tb{23}, 217-228 (2008).}
\bibitem{Nanofotonics1}{V. M. Shalaev and S. Kawata, \textit{Nanophotonics with surface Plasmons}. Elsevier (2007).}
\bibitem{Nanophotonics2}{P. G. Kik and M. L. Brongersma, \textit{Surface Plasmon Nanophotonics}, Springer Series in Optical Sciences, \tb{131}, 1-9, 2007. }
\bibitem{Stockman1}{K. F. MacDonald, Z. L. S\'amson, M. I. Stockman and N. I. Zheludev, \textit{Ultrafast active plasmonics}, Nature Photonics \tb{3}, 55 - 58 (2009)  }
\bibitem{biosensing}{J. N. Anker, W. P. Hall, O. Lyandres, N. C. Shah, J. Zhao and R. P. Van Duyne, \textit{Biosensing with plasmonic nanosensors}, Nature Materials,\tb{7}, 442-453 (2008). }
\bibitem{catalisis1}{X. Zhang, X. Ke, A. Du and H. Zhu,  \textit{Plasmonic nanostructures to enhance catalytic performance of zeolites under visible light}, Scientific Reports \tb{4}, 3805 (2014).}
\bibitem{catalisis2}{S. Linic,  P. Christopher and D.B. Ingram, \textit{Plasmonic-metal nanostructures for efficient conversion of solar to chemical energy}, Nature Materials, \tb{10}, 911-921 (2011)}.
\bibitem{Townsend}{E. Townsend and G. W. Bryant,  \textit{Plasmonic Properties of Metallic Nanoparticles: The Effects of Size Quantization},  Nano. Lett., \tb{12}, 429-434 (2012).}
\bibitem{Willets}{ K. A. Willets and R. P. Van Duyne,\textit{Localized surface plasmon resonance spectroscopy and sensing}, Annu. Rev. Phys. Chem. \textit{58}, 267–297 (2007).}
\bibitem{Juluri}{B. K. Juluri, Y. B. Zheng,  D. Ahmed, L. Jensen and T. J. Huang, T. J.\textit{Effects of geometry and composition on charge-induced plasmonic shifts in gold nanoparticles}, J. Phys. Chem. C \tb{112}, 7309–7317 (2008).}
\bibitem {Juan}{J. C. Arias and A. S. Camacho, \textit{Surface Plasmon Resonance of a Few Particles Linear Arrays}, JEMAA \tb{3}, 11 (2011).}
\bibitem {Carsten}{Carsten Sonnichsen \textit{Plasmons in metal nanostructures}, Ph. D Dissertation, Munich University, 2001.}
\bibitem {Wei}{ Q. H. Wei,K.-H. Su, S. Durant, and X. Zhang, \textit{Plasmon Resonance of Finite One-Dimensional Au Nanoparticle Chains}, Nano. Lett, \tb{4}, 1067 (2004).}
\bibitem {Murray}{W. A. Murray and W. L. Barnes, \textit{Plasmonics Materials}, Adv. Mat. \tb{19}, 3771 (2007)}.
\bibitem{Mie}{ G. Mie, \textit{Beitrage zur Optik truber Medien, speziell kolloidaler Metalloesunge} Ann. Phys. \tb{ 25}, 377 (1908).}
\bibitem{JandC}{P. B. Johnson and R. W. Christy, \textit{Optical Constants of the Noble Metals} Phys. Rev. B. \tb{6}, 12, 4370 (1975)} 
\bibitem{Garcia}{F. J. Garcia de Abajo, \textit{Plasmons go quantum}, \textit{Nature}, \tb{483}, 417, (2012).} 
\bibitem{Monreal}{R. C. Monreal, T. J. Antosiewicz and S. P. Apell, \textit{Competition between surface screening and size quantization for surface plasmons in nanoparticles}, New Journal of Physics \tb{15}, 083044 (2013)}
\bibitem{Scholl}{J. Scholl, A. L. Koh and J. Dionne.\textit{Quantum plasmon resonances of individual metallic nanoparticles}, Nature, \tb{483}, 421 (2012).}
\bibitem{Genzel}{L. Genzel and T. P.  Martin, \textit{Dielectric Function and Plasma Resonances of Small Metal Particles}, Z. Phys. B \tb{21}, 339 (1975).}
\bibitem{parametros}{In all our calculations, we use the standard set of parameters for silver taken from ref: \cite{Scholl,Carsten,JandC} $\epsilon_{\infty}=3.66$, $\omega_p=9.01$eV,  and $\gamma(R)=\gamma_{bulk}+\fr{A v_F}{R}$, with A=0.25 and $v_F=1.4\times10^6(m/s)$. }
\bibitem{Nordlander}{C. E. Talley, J. B. Jackson, C. Oubre, N. K. Grady, C. W. Hollars, S. M. Lane, T. R. Huser, P. Nordlander and N. J. Halas, \textit{Surface-Enhanced Raman Scattering from Individual Au Nanoparticles and Nanoparticle Dimer Substrates}, Nano Lett. \tb{5} (8), 1569-1574 (2005).}
\bibitem{Arfken}{G. B. Arfken and H. J. Weber, \textit{Mathematical Methods for Physicists}, Harcourt/Academic Press (2001).}
\bibitem{Yi}{Y. He and T. Zeng, \textit{First principles study and model of dielectric functions of silver nanopartticles}, J. Phys. Chem. C, \tb{ 114}, 18023-18030 (2010). }
\bibitem{Wood}{D. Wood, \textit{Classical Size Dependence of the Work Function of Small Metallic Spheres}, Phys. Rev. Lett. \tb{46}, 11 (1981). }
\bibitem{constantes}{We use $f=0.01$ and $\epsilon_m=2.25$ corresponding to a nanosphere immersed in a glass matrix.}
\end{thebibliography}


\end{document}